\documentclass[pop,aps,floats,preprint,graphics,graphicx,amsmath,amssymb,superscriptaddress,showpacs,pre]{revtex4-1}
\usepackage{graphics,graphicx}
\begin{document}
\title{Finite system size effects on Drift Wave stability}

\author{F. Militello} 
\affiliation{EURATOM/CCFE Fusion Association, Culham Science Centre, Abingdon, Oxon, OX14 3DB, UK}
\author{M. Ottaviani}
\affiliation{CEA, IRFM, F-13108 Saint-Paul-lez-Durance, France }
\author{A. Wynn} 
\affiliation{Imperial College of Science, Technology and Medicine, London, UK}

\begin{abstract}

Unstable electrostatic resistive modes, driven by density gradients, are identified in a bounded sheared slab. The boundary conditions play a crucial role and are shown to change the nature of the problem, which is related to so called "universal" mode. The dispersion relation and the structure of the eigenmodes of the instability are derived and are shown to depend on a limited set of dimensionless parameters. The occurrence and possible impact of these modes on numerical simulations and actual plasmas are discussed.   

\end{abstract}

\maketitle

\section{INTRODUCTION}

Density gradients are a necessary feature of confined plasmas and are also common in natural systems. The inhomogeneity is a source of free energy, which can amplify naturally occurring plasma oscillations and lead to turbulence and transport of particles, energy and momentum. It is therefore no surprise that density driven instabilities, i.e. unstable drift-waves, were the subject of intense theoretical research in plasma physics since its very beginning. 

The controversial story of the so called "universal" instability started with the pioneering work by Krall and Rosenbluth \cite{Krall1962}, which first identified it in a slab with a shearless magnetic field and in an electrostatic collisionless regime (the drive was electron wave-particle resonances coupled with the density gradient). The same authors introduced in a later paper \cite{Krall1965} the effect of the magnetic shear, which turned out to be a very efficient mechanism to stabilize the mode in experimental conditions. In the following years, it was suggested that the instability could still manifest in a sheared geometry as a transient effect \cite{Coppi1966}, i.e. not as a normal eigenmode. These so called convective modes were described as wave packets which initially grow in amplitude when they are in the region of local instability (close to the resonant surface), but eventually decay as they move in the stable region, coherently with the overall stability of the system.

Coming back to the normal eigenmodes, the mathematical framework of the problem was significantly complicated by the magnetic shear \cite{Krall1965}, since the derivation of the dispersion relation now required the solution of a differential equation (the magnetic shear prevents the use of Fourier transforms in the cross filed direction). This implies that proper boundary conditions must be imposed in order to find the correct result. In \cite{Krall1965}, and in all the following collisionless works, the ion Landau damping was not explicitly treated but it was assumed to stabilize the mode far away from its resonant surface. Indeed, the Landau damping is more effective for modes with a large parallel wave number, which is proportional to the cross field distance from the rational surface. The boundary conditions for the "universal" instability must therefore allow a smooth connection between the region where the Landau damping is neglected and the region where it is important. It is with this argument that Perlstein and Berk \cite{Perlstein1969} challenged the results in \cite{Krall1965}, where the eigenmodes were assumed to decay to zero sufficiently far away from the resonant surface (i.e. at the \textit{entrance} of the Landau damping region). The authors of \cite{Perlstein1969} sustained that the correct boundary conditions should allow for "outgoing waves" which would transport their energy outwards where it would eventually be dissipated by wave-particle interactions. This simple correction led to a new dispersion relation predicting unstable normal eigenmodes (more unstable than the convective modes), thus resurrecting the "universal" instability. 

Ten years later, two independent groups \cite{Ross1978,Tsang1978} came to the conclusion that the electron dynamics was not properly treated in \cite{Perlstein1969} (the principal part of the plasma dispersion function was not included, an inheritance from \cite{Krall1965}). With "outgoing waves" and the complete electron response, under no condition normal eigenmodes could be found in the sheared slab. These results led to the apparent conclusion that the "universal" instability was not so universal after all.

With the realization that the collisionless "universal" mode is stable in sheared magnetic fields, the only piece of physics missing was the effect of the collisions, which was studied in \cite{Guzdar1978,Chen1979}. This was done on the basis that, in drift wave dynamics, dissipative effects in the collisional regime play a destabilizing role similar to resonant electron-wave interactions in the collisionless regime \cite{Horton1999}. In these papers the approach is completely fluid (no wave-particle interactions), an "outgoing wave" boundary condition is used and the problem was solved with both perturbative techniques (separating the resistive layer from the outer solution) and WKB methods. The surprising result was that the resistivity increases even further the stability of the system.

However, other effects such as toroidicity \cite{Taylor1976,Chen1979,Chen1980} or coupling to trapped electrons \cite{Liu1976} can overcome the magnetic shear stabilization and produce linearly unstable modes. In addition, drift waves can be metastable and sustain turbulence through non linear coupling \cite{Scott1990a,Scott1990b}.  

The work that we present is an ideal continuation of this line of research. In particular, we describe how wave reflection in a finite size system can destabilize the resistive "universal" instability. Differently from the standard calculations, the boundary conditions that we use could be interpreted as an overlapping of "outgoing" and "ingoing" waves which create an unstable standing wave in the plasma. Our analysis is performed in the framework of a linearised reduced two fluid model in which finite resistivity is included and wave particle interactions are neglected. Both our model and some of the analytical techniques that we use have similarities with those employed in \cite{Guzdar1978,Chen1979}, although the final conclusions are opposite (because of different boundary conditions).

The most straightforward application of our results is to numerical simulations, which are intrinsically bounded systems. Also periodicity in the cross field direction, a common boundary condition in codes, would entrap energy and produce the unstable standing wave that we describe below. In addition, our boundary conditions might also be compatible with the energy exchange that occurs between coupled resonant surfaces in toroidal geometry \cite{Taylor1976,Chen1980}, so that the results of our simple slab analysis could have counterparts in realistic plasmas.      

\section{Model and equations} 
 
We investigate a reduced electrostatic two-fluid model, with isothermal electrons, cold ions and valid for small but finite $\beta$. Our equations were originally derived by Hazeltine \textit{et al.} \cite{Hazeltine1985} and are consistent with those employed in previous studies of the "universal" instability \cite{Ross1978,Guzdar1978,Chen1979}. The dimensionless equations evolve the plasma density, $n$, the plasma vorticity, $U$, which is related to the electrostatic potential, $\phi$, and the parallel ion velocity, $v$. they are:
\begin{eqnarray}
\label{1} &\partial U/\partial t& +[\phi,U]=[J,\psi_{eq}],\\
\label{3} &\partial n/ \partial t&+[\phi,n]=\rho^{2}[J,\psi_{eq}]-\beta[v,\psi_{eq}],\\
\label{4} &\partial v/ \partial t&+[\phi,v]=-[n,\psi_{eq}],\\
\label{5} &&[\phi,\psi_{eq}]=[n,\psi_{eq}]-\eta(J-J_{eq}),\\
\label{6} &U&=\nabla^2 \phi.
\end{eqnarray}
Here $J$ is the parallel current density ($J_{eq}$ is assumed to be constant) and $[f,g]\equiv \partial_x f\partial_y g -\partial_y f\partial_x g$. Note that $[f,\psi_{eq}]\equiv \nabla_\parallel$ represents the parallel gradient since $\psi$ is the poloidal magnetic flux. The model contains three dimensionless parameters, which are the normalized resistivity, $\eta=1/S$, where $S$ is the Lundquist number, the plasma $\beta \equiv 4\pi p/B^2$ (note the unusual coefficient $4$ instead of $8$) and $\rho\equiv \rho_s/L_s$, which is the ion Larmor radius calculated with the electron temperature (which we call $\rho_s$) normalized to an equilibrium length scale (here the magnetic shear length, $L_s$). The problem is solved in a slab domain which extends in the radial direction, $x$, along which the equilibrium density changes, and a transverse ("poloidal") direction $y$, which is assumed to be periodic. Both coordinates are normalized with respect to $L_s$. The time is normalized with respect to the Alfven time, $\tau_A \equiv v_A/L_s$, with $v_A \equiv B/\sqrt{4\pi n_0 m_i}$ ($n_0$ is a characteristic density and $m_i$ is the ion mass). More details on the normalizations and on the physics described by the different terms can be found in \cite{Grasso2002,Militello2008}. 

Note that our model is an extension of the well known Hasegawa-Wakatani equations \cite{Hasegawa1982}, that are retrieved in the limit $\beta=0$. In other words, with respect to \cite{Hasegawa1982}, we treat the parallel compressibility in the density equation in a more complete way, allowing the coupling with the parallel ion velocity.

Assuming that the perturbations can be expressed in the form, $\phi(x,y,t)=\phi(x)exp(\gamma t+ik_yy)$, and linearizing the equations we obtain:
\begin{equation}
\label{7} \phi''+Q(x,\hat{\gamma})\phi=0,
\end{equation}      
where: 
\begin{equation}
\label{9} Q(x,\hat{\gamma}) \equiv -k_y^2 - \frac{x^2}{\hat{\eta}\hat{\gamma}+\rho^2x^2}\left(1+i\frac{\hat{\omega_*}}{\hat{\gamma}}+\frac{\beta}{\hat{\gamma}^2}x^2\right),
\end{equation}
and the prime symbol represents derivation with respect to $x$. In the last equation, we have introduced the electron diamagnetic frequency (normalized to the Alfven frequency, $\tau_A^{-1}$), $\omega_*\equiv (\rho_s/L_n)k_yc_s\tau_A$, with $c_s\equiv \sqrt{T_e/m_i}$. This quantity describes the equilibrium density gradient (assumed to be locally constant) through the length scale $L_n\equiv L_s n_{eq}/n'_{eq}$ (remember that $'$ represents derivation with respect to the \textit{normalized} radial coordinate). The normalized diamagnetic frequency is another dimensionless parameter, which controls the dynamics of the system together with the other three introduced above. Note also that all the quantities with a hat symbol are further normalized with respect to $k_y$. This notation will be dropped in the following but $\eta$, $\gamma$ and $\omega_*$ should always be assumed thus normalized unless otherwise stated.

The magnetic field is assumed to be sheared so that the equilibrium component of $B_y$ depends on $x$ and it vanishes at $x=0$ [$B_y(x)=\psi_{eq}'\approx x$ around the resonant surface, using a Taylor expansion], while $B_z$ is a constant. This also implies that the parallel wave number of the perturbation is given by: $k_\parallel = k_yx$ since $ik_\parallel=\textbf{B}_{eq}\cdot \nabla$ and $z$ is taken to be an ignorable direction.

The problem \ref{7}-\ref{9} is completely defined only when the boundary conditions are imposed, which, as discussed in the introduction, can strongly affect the final result. While the "outgoing wave" boundary conditions would require a solution that carries energy outwards, we impose that the wave is completely reflected at a finite distance $L_x$ from either side of the resonant surface, which is equivalent to setting $\phi(\pm L_x)=0$. Therefore, the energy does not leak out of the system, not differently from a configurations with periodic boundary conditions in $x$ (which would lead to very similar conclusions as discussed in the Appendix). 

We solved Eqs.\ref{7}-\ref{9} in various limits in order to clarify the effect of the different terms. In some cases, it is possible to obtain an exact solution of the problem, while in others, the solution can only be found by making use of small parameters and applying perturbative techniques. 

For convenience, and similarly to \cite{Guzdar1978,Chen1979}, the potential $Q$ is rewritten in the following form:
\begin{equation}
\label{35b} Q(x,\gamma) = \delta-\epsilon^2 x^2 -\frac{\Lambda}{x^2+x_R^2},
\end{equation}
with:
\begin{eqnarray}
\label{35c} \delta &\equiv& -k_y^2-\frac{\gamma+i\omega_*}{\rho^2\gamma}+\frac{\eta\beta}{\rho^4\gamma},\\
\label{35d} \epsilon^2 &\equiv& \frac{\beta}{\rho^2\gamma^2},\\
\label{35e} \Lambda &\equiv& -\frac{\eta(\gamma+i\omega_*)}{\rho^4}+\frac{\eta^2\beta}{\rho^6},\\
\label{35f} x_R^2 &\equiv & \frac{\eta\gamma}{\rho^2}. 
\end{eqnarray}

\subsection{Dispersion relation in local approximation}

To introduce the problem, we first study the dispersion relation in the limit of zero magnetic shear. This can be done by assuming a constant parallel wavelength, $k_\parallel =k_{\parallel,0}$. This approximation removes all the radial dependencies from the equilibrium coefficients of the equations, thus allowing a Fourier transformation also in the $x$ direction. The problem becomes scalar and trivial to solve once $x$ is replaced by $k_{\parallel,0}/k_y$ and $\phi''$ by $-k_x^2\phi$ in Eqs.\ref{7} and \ref{9}. The dispersion relation is:
\begin{equation}
\label{9a} \gamma^3\left(\eta k_y^2 \frac{k_\perp^2}{k_\parallel^2}\right) +\gamma^2(1+\rho^2k_\perp^2) + \gamma(i\omega_*) +\beta \frac{k_\parallel^2}{k_y^2}=0,
\end{equation} 
where $k_\perp^2 \equiv k_x^2+k_y^2$. In order to find a simple solution of this equation, we assume $\eta \ll 1$, and we expand in this small parameter, so that $\gamma=\gamma_0+\gamma_1$, with $\gamma_1$ of order $\eta$. This gives:
\begin{eqnarray}
\label{9b} \gamma_0 && \approx -\frac{i\omega_*}{1+\rho^2 k_\perp^2}\left(\frac{1}{2}\pm\frac{1}{2}\sqrt{1+4\beta\frac{1+\rho^2k_\perp^2}{\omega_*^2}\frac{k_\parallel^2}{k_y^2}}\right), \\
\label{9c} \gamma_1 && \approx \pm\eta \omega_*^2\frac{k_y^2}{k_\parallel^2} \frac{k_\perp^2}{(1+\rho^2 k_\perp^2)^3}\left(\frac{1}{2}\pm\frac{1}{2}\sqrt{1+4\beta\frac{1+\rho^2k_\perp^2}{\omega_*^2}\frac{k_\parallel^2}{k_y^2}}\right)^3\left(\sqrt{1+4\beta\frac{1+\rho^2k_\perp^2}{\omega_*^2}\frac{k_\parallel^2}{k_y^2}}\right)^{-1},
\end{eqnarray}
where the solution with the plus sign in Eq.\ref{9c} describes a resistive instability driven by the density gradient. Note that in the case $\beta=\eta=0$ we recover the standard drift waves, while for $\omega_*=\eta=0$ we find sound waves. Equation \ref{9a} has a third solution, which is singular for $\eta\rightarrow 0$, but it is stable and therefore not interesting. The growth rate in Eq.\ref{9c} gives some insight on the nature of the instabilities we discuss in this paper. They are resistive (for $\eta=0$, $\gamma_1=0$) and driven by the density gradients (through $\omega_*$). The same features are observed also in the presence of magnetic shear, as the following Sections show.

\section{Exact results} \label{Exact}

\subsection{Case with: $\rho=\beta=0$} \label{SecIVA}

At low temperature $\rho$ and $\beta$ are small and they can be neglected in Eq.\ref{9} so that:
\begin{equation}
\label{11} Q_{\rho=\beta=0}(x,\gamma) = -k_y^2 - \frac{\gamma+i\omega_*}{\eta\gamma^2}x^2=\widehat{\delta}-\widehat{\epsilon}^2x^2,
\end{equation}   
where $\widehat{\delta}\equiv \delta-\Lambda/x_R^2$ and $\widehat{\epsilon}^2\equiv -\Lambda/x_R^4$ are evaluated at $\beta=0$.
In this limit, Eq.\ref{7} coupled with Eq.\ref{11} has an exact solution in terms of Whittaker functions \cite{AbramovitzBOOK}:
\begin{equation}
\label{12} \phi = C_1x^{-1/2}M_{\widehat{\kappa},1/4}\left(\widehat{\epsilon} x^2\right)+C_2x^{-1/2}W_{\widehat{\kappa},1/4}\left(\widehat{\epsilon} x^2\right),
\end{equation} 
where $\widehat{\kappa}=\frac{1}{4}\frac{\widehat{\delta}}{\widehat{\epsilon}}$, $C_1$ and $C_2$ are complex integration constants. 

Modes with even and odd parity with respect to $x$ are characterized by $\phi'(0)=0$ and $\phi(0)=0$, respectively. The first condition corresponds to:
\begin{equation}
\label{18} C_1=C_2\frac{2\sqrt{\pi}}{\Gamma\left(\frac{1}{4}-\widehat{\kappa}\right)},
\end{equation} 
while the second gives:
\begin{equation}
\label{18a} C_2=0.
\end{equation} 
Imposing also the reflecting boundary condition, we obtain the dispersion relation $M_{\widehat{\kappa},\pm1/4}(\widehat{\epsilon}L_x^2) = 0$, where the plus and minus signs are for odd and even modes, respectively. In the limit of small resistivity, $\widehat{\epsilon} \gg \widehat{\delta}$, this simplifies to: $J_{\pm 1/4}(-i\widehat{\epsilon}L_x^2/2)=0$. We use now approximate relations to find the zeros of the Bessel function \cite{AbramovitzBOOK}, obtaining the dispersion relation: $\widehat{\epsilon}\approx i 2 \pi(n\pm 1/8-1/4)/L_x^2$, with $n \in \mathbb{N}$ and the usual parity convention for the plus and minus sign. Using the definition of $\widehat{\epsilon}$, we find:
\begin{eqnarray}
\label{31}  && \gamma_r^2(\omega_*-\gamma_i)-\gamma_i^2(\omega_*+\gamma_i)=0, \\
\label{32} &&  \frac{\gamma_r(\gamma_r^2+\gamma_i^2+2\gamma_i\omega_*)}{\eta(\gamma_r^2+\gamma_i^2)^2}=-\frac{4\pi^2}{L_x^4}\left(n\pm\frac{1}{8}-\frac{1}{4}\right)^2, 
\end{eqnarray}
where we have used the fact that $\Re(\widehat{\epsilon})=0$ implies that $\Im(\widehat{\epsilon}^2)=0$ and therefore $\Im(\widehat{\epsilon})^2=-\Re(\widehat{\epsilon}^2)$. In the previous expressions, $\gamma_r\equiv \Re(\gamma)$ and $\gamma_i \equiv \Im(\gamma)$ are the growth rate of the mode and its rotation frequency.

Equation \ref{31} shows that, in the spectrum, the unstable eigenvalues are located on a curve that does not depend on the resistivity, but only on the diamagnetic frequency. Consequently, also the growth rate of the fastest growing mode depends only on $\omega_*$: $\gamma_{r,max}\approx 0.3\omega_*$. Note also that both even and odd modes lie on the same curve, described by Eq.\ref{31}. Finally, it is important to remark that the complex frequency of the single modes retains a dependency on $\eta$ and their growth rate vanishes in the zero resistivity limit (when all the modes collapse on the point $[\gamma_r,\gamma_i]=[0,-\omega_*]$) or in the infinite resistivity limit (when all the modes collapse on the point $[\gamma_r,\gamma_i]=[0,0]$).

\subsection{Case with: $\eta=0$} \label{SecIVB}
 
For $\eta=0$ Eqs.\ref{35b}-\ref{35f} yield a quadratic potential $Q(x,\gamma)$ that is structurally similar to the one discussed in the previous section (i.e Eq.\ref{11}). Hence, the procedure described above can be straightforwardly extended to the ideal case ($\eta=0$), valid for high temperature collisionless plasmas. In this regime, the condition $\Im(\epsilon^2)=0$ translates into $\gamma_i\gamma_r=0$ so that the mode either rotates, but does not grow, or vice versa. The condition on $\Re(\epsilon^2)$, analogous to Eq.\ref{32}, would give:
\begin{equation}
\label{35} \Re(\epsilon^2) = \frac{\beta}{\rho^2}\frac{\gamma_r^2-\gamma_i^2}{(\gamma_r^2+\gamma_i^2)^2} =-\frac{4\pi^2}{L_x^4}\left(n-\pm \frac{1}{8}+\frac{1}{4}\right)^2
\end{equation}
As the right-hand side of the previous equation is negative defined, the only acceptable solution is $|\gamma_r|<|\gamma_i|$, which leads to $\gamma_r=0$ and implies no growing modes. This result implies that in the fluid limit the resistivity is a key ingredient to destabilize the modes discussed in this paper. A proper investigation of the effect of finite system size in collisionless regimes would require a kinetic treatment which is beyond the scope of the present study.

\section{Perturbative approach} \label{Perturbative}
 
As suggested by the previous Sections, a finite resistivity is essential to produce unstable modes in finite size fluid systems. This contrasts with earlier analytic results \cite{Guzdar1978,Chen1979}, which used outgoing wave boundary conditions and predicted stability and even a damping due to $\eta$. We now want to asses the properties of the unstable modes when all the physics is included in the problem. Unfortunately, in this case an exact solution of Eq.\ref{7}-\ref{9} is not available. On the other hand, an asymptotic matching approach, similar to the one used in \cite{Guzdar1978,Chen1979}, provides an approximate solution.

\subsection{Outer region}
We start by studying the outer region, located at $x\gg x_R$. In this case, $Q$ reduces to:
\begin{equation}
\label{35g} Q_{out}\equiv \delta-\epsilon^2 x^2 -\frac{\Lambda}{x^2},
\end{equation}
the solution of which can be expressed as:
\begin{equation}
\label{35h} \phi_{out} = C_1x^{-1/2}M_{\kappa,\mu}\left(\epsilon x^2\right)+C_2x^{-1/2}W_{\kappa,\mu}\left(\epsilon x^2\right),
\end{equation} 
(compare with Eq.\ref{12}), with the following indexes: $\kappa\equiv \frac{1}{4}\frac{\delta}{\epsilon}$, $\mu=-\sqrt{1+4\Lambda}/4$, $C_1$ and $C_2$ are the constant of integration.  
 
\subsection{Inner Region} 
 
The inner region is characterized by $x\sim x_R \ll 1$. In this limit, we have:
\begin{equation}
\label{48a} Q_{in} = -\frac{\Lambda}{x^2+x_R^2},
\end{equation}
and the solution of Eq.\ref{7} can be expressed in terms of Legendre functions:
\begin{equation}
\label{49} \phi_{in} \approx C_3 \sqrt{x^2+x_R^2}P^1_\nu\left(\frac{x}{ix_R}\right)+C_4 \sqrt{x^2+x_R^2}Q^1_\nu\left(\frac{x}{ix_R}\right),
\end{equation} 
where $C_3$ and $C_4$ are constants of integration and $\nu\equiv 1/2(\sqrt{1+4\Lambda}-1)=-(1/2+2\mu)$. The Legendre functions are multivalued on on the real axis between $-1\leq \Re(x/ix_R) \leq 1$, so here they have a branch cut. This can be taken into account by taking $\Im[(ix_R)^{-1}]<0$ (valid for unstable modes, $\Re(x_R)>0$) and $x>0$, i.e. we go to zero from the third or fourth quadrant. At $x=0$, this leads to:
\begin{eqnarray}
\label{50} \phi_{in}(0) &=& \sqrt{x_R^2}\frac{\Gamma\left(\frac{1}{2}\nu+1\right)}{\Gamma\left(\frac{1}{2}\nu+\frac{1}{2}\right)}\left\{\left(-iC_3\frac{2}{\sqrt{\pi}}+C_4\sqrt{\pi}\right)\sin(\pi\nu/2)-iC_4\sqrt{\pi}\cos(\pi\nu/2)\right\},\\
\label{51} \phi'_{in}(0) &=& \sqrt{x_R^2}\frac{\Gamma\left(\frac{1}{2}\nu+\frac{3}{2}\right)}{\Gamma\left(\frac{1}{2}\nu\right)}\left\{\left(iC_3\frac{4}{\sqrt{\pi}}-C_42\sqrt{\pi}\right)\cos(\pi\nu/2)-iC_42\sqrt{\pi}\sin(\pi\nu/2)\right\}.
\end{eqnarray}
which implies that the odd modes [$\phi_{in}(0)=0$] have:
\begin{equation}
\label{52} C_4=-C_3\frac{2}{\pi}\sin(\nu\pi/2)e^{-i\nu\pi/2}.
\end{equation} 
Similarly, for the even modes [$\phi_{in}'(0)=0$]:
\begin{equation}
\label{52b} C_4=iC_3\frac{2}{\pi}\cos(\nu\pi/2)e^{-i\nu\pi/2}.
\end{equation} 

\subsection{Matching and dispersion relation} 
 
Let's now match the inner and the outer solutions. To do this, we first take the small $x$ limit of the outer solution:
\begin{equation}
\label{53} \phi_{out}(x)\approx \epsilon^{\frac{1}{2}+\mu}\left[C_1+C_2\frac{\Gamma(-2\mu)}{\Gamma\left(\frac{1}{2}-\mu-\kappa\right)}\right]x^{\frac{1}{2}+2\mu}+C_2\frac{\epsilon^{\frac{1}{2}-\mu}}{2\mu}\frac{\Gamma(1+2\mu)}{\Gamma\left(\frac{1}{2}+\mu-\kappa\right)}x^{\frac{1}{2}-2\mu},
\end{equation}
The large $x$ limit of the inner solution is:
\begin{eqnarray}
\label{54} \phi_{in}(x) \approx &&C_3 \frac{2^\nu\Gamma\left(\nu+\frac{1}{2}\right)}{\sqrt{\pi}\Gamma(\nu)}\left(\frac{1}{ix_R}\right)^{\nu}x^{\nu+1}+\nonumber \\
&& \left[C_3\frac{\Gamma\left(-\nu-\frac{1}{2}\right)}{2^{\nu+1}\sqrt{\pi}\Gamma(-\nu-1)}-C_4\frac{\sqrt{\pi}}{2^{\nu+1}}\frac{\Gamma(\nu+2)}{\Gamma\left(\nu+\frac{3}{2}\right)}\right]\left(\frac{1}{ix_R}\right)^{-\nu-1}x^{-\nu}
\end{eqnarray} 
Reminding that $-\nu= 1/2+2\mu$, we match the first term of $\phi_{in}$ with the second of $\phi_{out}$ and the second term of $\phi_{in}$ with the first of $\phi_{out}$.

This gives:
\begin{eqnarray}
\label{55} &&C_3 \frac{2^\nu\Gamma\left(\nu+\frac{1}{2}\right)}{\sqrt{\pi}\Gamma(\nu)}\left(\frac{1}{ix_R}\right)^{\nu}=-C_2\frac{\epsilon^{\frac{3}{4}+\frac{\nu}{2}}}{1/2+\nu}\frac{\Gamma\left(\frac{1}{2}-\nu\right)}{\Gamma\left(\frac{1}{4}-\frac{\nu}{2}-\kappa\right)},\\
\label{56}  &&\left[C_3\frac{\Gamma\left(-\nu-\frac{1}{2}\right)}{2^{\nu+1}\sqrt{\pi}\Gamma(-\nu-1)}-C_4\frac{\sqrt{\pi}}{2^{\nu+1}}\frac{\Gamma(\nu+2)}{\Gamma\left(\nu+\frac{3}{2}\right)}\right]\left(\frac{1}{ix_R}\right)^{-\nu-1} = \epsilon^{\frac{1}{4}-\frac{\nu}{2}}\left[C_1+C_2\frac{\Gamma\left(\frac{1}{2}+\nu\right)}{\Gamma\left(\frac{3}{4}+\frac{\nu}{2}-\kappa\right)}\right].
\end{eqnarray}
From the first equation, we find:
\begin{equation}
\label{57} C_2 = -C_3 \left(\nu+\frac{1}{2}\right)\epsilon^{-\frac{3}{4}-\frac{\nu}{2}}\frac{2^{\nu}\Gamma\left(\nu+\frac{1}{2}\right)\Gamma\left(\frac{1}{4}-\frac{\nu}{2}-\kappa\right)}{\sqrt{\pi}\Gamma(\nu)\Gamma\left(\frac{1}{2}-\nu\right)}\left(\frac{1}{ix_R}\right)^{\nu}
\end{equation}

If we assume a finite box size, we have to impose the condition:
\begin{equation}
\label{59} C_1=-C_2\frac{W_{\kappa,\mu}(\epsilon L_x^2)}{M_{\kappa,\mu}(\epsilon L_x^2)},
\end{equation}
which leads to:
\begin{eqnarray}
\label{60} \frac{\Gamma\left(\frac{1}{4}-\frac{\nu}{2}-\kappa\right)}{\Gamma\left(\frac{3}{4}+\frac{\nu}{2}-\kappa\right)} &=&-\frac{\Gamma(\nu)\Gamma\left(\frac{1}{2}-\nu\right)}{\Gamma\left(\nu+\frac{1}{2}\right)\Gamma\left(\nu+\frac{3}{2}\right)}\left[\frac{\Gamma\left(-\nu-\frac{1}{2}\right)}{\Gamma(-\nu-1)}-\frac{C_4}{C_3}\frac{\pi\Gamma(\nu+2)}{\Gamma\left(\nu+\frac{3}{2}\right)}\right]\left(\frac{i\sqrt{\epsilon}x_R}{2}\right)^{2\nu+1} +\nonumber \\
&&+ \frac{W_{\kappa,\mu}(\epsilon L_x^2)}{M_{\kappa,\mu}(\epsilon L_x^2)}\frac{\Gamma\left(\frac{1}{4}-\frac{\nu}{2}-\kappa\right)}{\Gamma\left(\frac{1}{2}+\nu\right)}. 
\end{eqnarray}
Note that $W_{\kappa,\mu}(\epsilon L_x^2)/M_{\kappa,\mu}(\epsilon L_x^2)$ goes to zero when $L_x$ goes to infinity so that in this limit we correctly recover the dispersion relation in \cite{Guzdar1978,Chen1979}. It is convenient to express $W_{\kappa,\mu}$ in terms of $M_{\kappa,\pm\mu}$, which gives a more compact version of the new dispersion relation:
\begin{eqnarray}
\label{61} \frac{M_{\kappa,-\mu}(\epsilon L_x^2)}{M_{\kappa,\mu}(\epsilon L_x^2)} &=&\frac{\Gamma(\nu)\Gamma\left(\frac{1}{2}-\nu\right)}{\Gamma\left(-\nu-\frac{1}{2}\right)\Gamma\left(\nu+\frac{3}{2}\right)}\left[\frac{\Gamma\left(-\nu-\frac{1}{2}\right)}{\Gamma(-\nu-1)}-\frac{C_4}{C_3}\frac{\pi\Gamma(\nu+2)}{\Gamma\left(\nu+\frac{3}{2}\right)}\right]\left(\frac{i\sqrt{\epsilon}x_R}{2}\right)^{2\nu+1}.
\end{eqnarray}

\subsection{Dimensionless parameters and relevant regimes} \label{dimpar}

Only a limited number of combinations of dimensionless parameters appear in the dispersion relation, Eq.\ref{61}. They are:
\begin{eqnarray}
\label{61a} \nu &=& i \left(\frac{\sqrt{\eta \omega_*}}{\rho^2}\right)^2(\overline{\gamma} +1), \nonumber\\
\label{61b} \kappa &=& -i\frac{[1+\overline{\gamma}(1+\rho^2k_y^2)]}{4}\left(\frac{\rho\sqrt{\beta}}{\omega_*}\right)^{-1}, \nonumber \\
\label{61c} \mu &=& -\frac{1}{4}-i\left(\frac{\sqrt{\eta \omega_*}}{\rho^2}\right)^2\frac{(\overline{\gamma} +1)}{2}, \nonumber \\
\label{61d} \epsilon L_x^2 &=& -i\overline{\gamma}^{-1}\left(\frac{\rho \sqrt{\beta}}{\omega_*}\right)\left(\frac{L_x}{\rho}\right)^2, \nonumber \\
\label{61e} \sqrt{\epsilon}x_R &=& \left(\frac{\sqrt{\eta\omega_*}}{\rho^2}\right)\left(\frac{\rho\sqrt{\beta}}{\omega_*}\right)^{1/2}.
\end{eqnarray}
with $\overline{\gamma} = \gamma/(i\omega_*)$ the rescaled complex frequency. In an abstract form, we can therefore write Eq.\ref{61} as $f(\overline{\gamma}, \sqrt{\eta\omega_*}/\rho^2,\rho k_y, L_x/\rho, \rho\sqrt{\beta}/\omega_*)=0$ where only four parameters control the problem. While some of them have a straightforward interpretation, it is useful to point out that $\rho\sqrt{\beta}/\omega_*=L_n/L_s$ (in the following we use only the latter definition). These new parameters determine the properties of the modes even when no perturbative expansion is involved, as can be seen by properly renormalizing Eqs.\ref{7} and \ref{9}. We note, for example, that the conclusions in Section \ref{SecIVA} are valid when $L_x/\rho \ll L_s/L_n$ together with $L_x/\rho \ll (\sqrt{\eta\omega_*}/\rho^2) \overline{\gamma}^{1/2}$. Similarly, the regime described in Section \ref{SecIVB} corresponds to the opposite limit: $L_x/\rho \gg (\sqrt{\eta\omega_*}/\rho^2) \overline{\gamma}^{1/2}$. In addition, when $L_x/\rho \gg \overline{\gamma}^{1/2}(L_s/L_n)^{1/2}$, we are able to reproduce the results discussed in \cite{Guzdar1978,Chen1979}. In the next Sections, we study other relevant regimes in which the complete dispersion relation, Eq.\ref{61}, has a transparent interpretation.  

\subsection{Small resistivity limit ($\sqrt{\eta\omega_*}/\rho^2\ll 1$)} \label{Small_res}

From Eq.\ref{61a} it is clear that in this limit also $|\nu|\ll 1$, which leads to several simplifications (we discuss solutions for which $|\overline{\gamma}|$ is of order unity or smaller) and that $\mu \approx -1/4$. We start by studying the even modes. In this case, Eq.\ref{61} becomes:
\begin{equation}
\label{62} \frac{M_{\kappa,1/4}\left(\epsilon L_x^2\right)}{M_{\kappa,-1/4}\left(\epsilon L_x^2\right)}+O(\nu)=-\frac{2}{\pi}\frac{h}{\sqrt{\nu}}+O[\nu^{1/2}\log(\nu)]
\end{equation}
where: $h\equiv -i^{3/2}(\overline{\gamma}+1)^{-1/2}(L_n/L_s)^{1/2}$. This can also be written as:
\begin{equation}
\label{63a} \frac{\sqrt{\overline{\gamma}}}{\overline{\gamma}+1}\frac{M\left(\frac{1}{4}-\kappa,\frac{1}{2},\epsilon L_x^2\right)}{M\left(\frac{3}{4}-\kappa,\frac{3}{2},\epsilon L_x^2\right)}=-i^{1/2}\frac{\pi}{2}\left(\frac{\sqrt{\eta\omega_*}}{\rho^2}\right)\frac{L_x}{\rho}.
\end{equation}
Note that while $M_{\kappa,\epsilon}(z)$ refers to Whittaker functions, the notation $M(a,b,z)$ is used for confluent hypergeometrics  \cite{AbramovitzBOOK}.

We discuss now the odd modes, which give the following dispersion relation:
\begin{equation}
\label{62a} -2\frac{M_{\kappa,1/4}\left(\epsilon L_x^2\right)}{M_{\kappa,-1/4}\left(\epsilon L_x^2\right)}+O(\nu)=-\pi h\nu^{3/2}+O[\nu^{5/2}\log(\nu)].
\end{equation}
In this case, the right hand side is higher order and we need to treat explicitly the order $\nu$ term at the left hand side, which is dominant. This observation allows us to determine the correct dispersion relation directly from Eq.\ref{61}, in which we can neglect the right hand side. The solution is obtained by evaluating the zeros of $M_{\kappa,-\mu}(\epsilon L_x^2) \sim M(1/2-\mu-\kappa,1-2\mu,\epsilon L_x^2)$ when $\nu$ is small. The zeros of a confluent hypergeometric are given by an approximated expression \cite{AbramovitzBOOK}:
\begin{equation}
\label{62b} M(a,b,x_0)=0 \rightarrow x_0 \approx \frac{\pi^2}{2}\frac{(n+b/2-3/4)^2}{b-2a}
\end{equation}
when $n \in \mathbb{N}$ is sufficiently large. Applying this formula to our problem, we find:
\begin{equation}
\label{62c} \delta L_x^2 \approx \pi^2\left(n+\frac{\nu}{2}\right)^2,
\end{equation}
which is independent from the parameter $L_n/L_s$. Note, however, that this solution is valid only if the order $\nu$ term on the left hand side of Eq.\ref{62a} is larger than the right hand side. This is true when $\pi\nu^{1/2} h \ll 1$, which corresponds to $\pi^2\sqrt{\eta\omega_*}/\rho^2\ll \left(L_s/L_n\right)^{1/2}$. If we remain in this limit, Eq.\ref{62c} is valid and it can be written as:
\begin{equation}
\label{62d} \gamma \approx -\frac{i\omega_*}{1+\rho^2k_{\perp,n}^2}+\frac{\pi^2\eta\omega_*^2}{L_x^2}\frac{n k_{\perp,n}^2}{(1+\rho^2k_{\perp,n}^2)^3},
\end{equation}
with $k^2_{\perp,n}\equiv k_y^2+n^2\pi^2/L_x^2$ and $n \in \mathbf{N}$. The maximum growth rate for the odd modes can be easily calculated from Eq.\ref{62d} and, in the limit $\rho k_y\ll 1$, it is $\Re(\gamma)_{max}\approx 0.39\omega_*(\sqrt{\eta\omega_*}/\rho^2)^2(\rho/L_x)$ with corresponding rotation frequency $\Im(\gamma)_{max}\approx -\omega_*/2$.

\subsubsection{$L_n/L_s \rightarrow 0$ limit for the even modes} \label{Small_res_even}

Equation \ref{63a} can be simplified if we assume a small $\rho\sqrt{\beta}/\omega_*=L_n/L_s$ limit, equivalent to performing a secondary expansion that decouples the sound waves and reduces the system to the Hasegawa-Wakatani model. When $L_n/L_s\ll 1$,  $|\kappa|$ becomes large, so that the confluent hypergeometric function transforms into a Bessel function \cite{AbramovitzBOOK}. Eq.\ref{63a} becomes:
\begin{equation}
\label{64} \frac{\sqrt{\overline{\gamma}(1+k_y^2\rho^2)+1}}{\overline{\gamma}+1}\frac{J_{-1/2}(\sqrt{\delta}L_x)}{J_{1/2}(\sqrt{\delta}L_x)}=i^{3/2}\frac{\pi}{2}\frac{\sqrt{\eta\omega_*}}{\rho^2},
\end{equation}
where the fraction of the Bessel functions above is a cotangent: $J_{-1/2}(\sqrt{\delta}L_x)/J_{1/2}(\sqrt{\delta}L_x) =\cot\left(i\frac{L_x}{\rho}\sqrt{1+\rho^2k_y^2+\frac{1}{\overline{\gamma}}}\right)$.
By noting that the dispersion relation is such that:
\begin{equation}
\label{65} F\left(\overline{\gamma},\rho k_y, \frac{L_x}{\rho}\right)=i^{3/2}\frac{\pi}{2}\frac{\sqrt{\eta\omega_*}}{\rho^2} \ll 1,
\end{equation}
we can expand the complex frequency using the small parameter $\sqrt{\eta\omega_*}/\rho^2$:
\begin{equation}
\label{66} F\left(\overline{\gamma}_0,\rho k_y, \frac{L_x}{\rho}\right)+\frac{dF}{d\overline{\gamma}}\left(\overline{\gamma}_0,\rho k_y, \frac{L_x}{\rho}\right)\overline{\gamma}_1\approx i^{3/2}\frac{\pi}{2}\frac{\sqrt{\eta\omega_*}}{\rho^2}.
\end{equation}
This gives $F(\overline{\gamma}_0)\approx 0$ the solution of which is: $\overline{\gamma}_0 \approx -(1+\rho^2k_{\perp n}^2)^{-1}$, where $k_{\perp n}^2\equiv k_y^2 +(\pi/L_x)^2(n+1/2)^2$ and $n \in \mathbb{N}_0$.

The equation $\left.\frac{dF}{d\overline{\gamma}}\right|_{\overline{\gamma}_0}\overline{\gamma}_1\approx i^{3/2}\frac{\pi}{2}\frac{\sqrt{\eta\omega_*}}{\rho^2}$ leads to:
\begin{equation}
\label{67} \overline{\gamma}_1 \approx -i^{3/2}\pi \frac{\sqrt{\eta \omega_*}}{\rho^2}\frac{\rho}{L_x}\frac{\rho^2k_{\perp n}^2}{(1+\rho^2k_{\perp n}^2)^{5/2}}, 
\end{equation}
so that the complex frequency has a positive growth rate corresponding to an instability:
\begin{equation}
\label{68} \gamma \approx -\frac{i\omega_*}{1+\rho^2k_{\perp n}^2}+\frac{\sqrt{2}\pi}{2}(1+i)\sqrt{\eta} \omega_*^{3/2}\frac{\rho}{L_x}\frac{k_{\perp n}^2}{(1+\rho^2k_{\perp n}^2)^{5/2}}.
\end{equation}
We can now determine the behaviour of the fastest growing mode by setting $\partial_n[k_{\perp n}^2(1+\rho^2k_{\perp n}^2)^{-5/2}]=0$, which gives $n_{max} = int\left[\frac{1}{\pi}\frac{L_x}{\rho}\sqrt{\frac{2}{3}-\rho^2k_y^2}-\frac{1}{2}\right]$, where the operator $int[\cdots]$ rounds up its argument to the closest integer. By replacing $n_{max}$ in Eq.\ref{68}, we find that the fastest growing mode has a growth rate $\Re(\gamma)_{max}\approx 0.41 \omega_*(\sqrt{\eta\omega_*}/\rho^2)(\rho/L_x)$ and its rotation frequency is $\Im(\gamma)_{max} \approx -(3/5)\omega_*$.

\subsubsection{Finite $L_n/L_s$ corrections}\label{FbT}

Despite our best efforts, we could not identify a limit for Eqs.\ref{63a} and \ref{62a} which contained $L_n/L_s$ effects and that could be expressed with a transparent analytical formulation. As a consequence, this regime was studied numerically with a thorough characterisation of the complex frequency as the dimensionless parameters are varied. The results of this investigation are reported in Section \ref{FbN}.

\section{Numerical Results} \label{Numerical}

In order to verify our analytic results, we solved the linear version of Eqs.\ref{1}-\ref{6} with a finite difference spectral code. Its output provided the full spectrum of the eigenvalues and eigenfunctions of the system, including sub-dominant instabilities and stable modes. In this Section we verify numerically the results obtained in Sec. \ref{Exact} and \ref{Perturbative} and we also determine how the modes behave at finite $L_n/L_s$ values, where no simple analytic limit was found. 

\subsection{small $\rho$ and $\beta$}

\begin{figure}
\includegraphics[height=8cm]{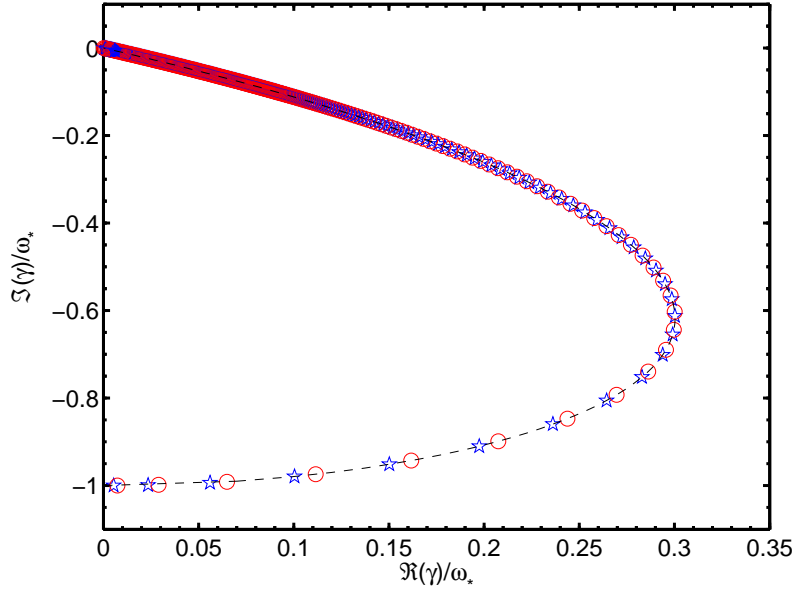} \caption{Comparison between numerical spectrum and theoretical predictions for $\rho=\beta=0$ and $\sqrt{\eta\omega_*}/L_x^2=0.0035$. The stars and circles represent modes, respectively with even and odd parity, which are calculated numerically. The mode number, $n$ gradually increases from the bottom to the top of the figure. The dashed line shows the theoretical prediction of Eq.\ref{31}.}
\label{fig1}
\end{figure}

\begin{figure}
\includegraphics[height=8cm]{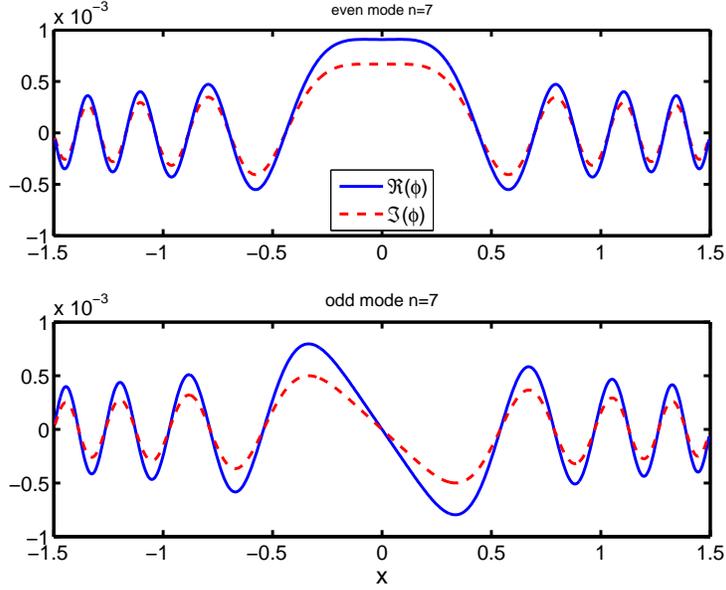} \caption{Mode structure of the electrostatic potential for an even and odd parity perturbation with $n=7$ for the parameters in Fig.\ref{fig1}. $L_x=1.5$ in this case.}
\label{fig2}
\end{figure}

Inspection of Eq.\ref{32} reveals that the growth rate of the modes depends only on the combination $(\sqrt{\eta\omega_*}/\rho^2)(\rho/L_x)^2$. Indeed, using the dimensionless complex frequency introduced in Sec.\ref{dimpar}, we have that the unstable branch of the dispersion relation is $\overline{\gamma}=i(1- \sqrt{1-4ia})/a$ with $a\equiv 4\pi^2(\eta\omega_*/L_x^4)(n\pm1/8-1/4)^2$. This dispersion relation perfectly matches the numerical spectrum, given in Fig.\ref{fig1}.
In Fig.\ref{fig2} we give an example of the odd and even parity eigenfunctions associated with the $n=7$ mode number. A similar oscillating structure of the perturbations is retrieved also in the more complicated regimes described in the following sections.

\subsection{Small $\sqrt{\eta\omega_*}/\rho^2$ regime}

\begin{figure}
\includegraphics[height=8cm]{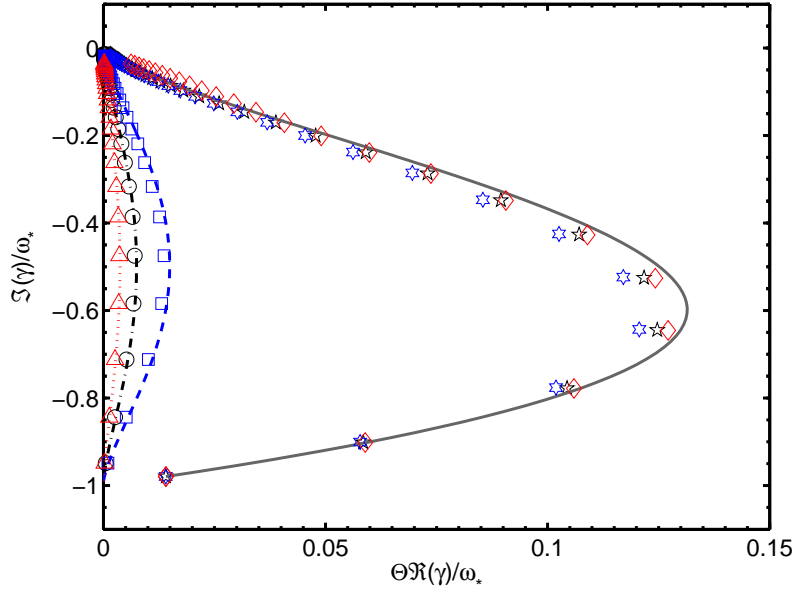} \caption{Normalized numerical spectrum of the modes. Even modes are shown as diamonds, stars and hexagrams for $\sqrt{\eta\omega_*}/\rho^2$ equals to 0.03, 0.06, 0.12. Odd modes are shown as triangles, circles and boxes for the same range of $\sqrt{\eta\omega_*}/\rho^2$. The solid line is the theoretical prediction Eq.\ref{69} for the even modes. The dashed, dot-dashed and dotted lines are the theoretical predictions for the odd modes given by Eq.\ref{70}. Note that in the horizontal axis, the growth rate is multiplied by the factor $\Theta$ (see text). $L_x/\rho=15$ and $\rho k_y = 0.1$ for all the cases shown.}
\label{fig3}
\end{figure}

In order to compare the numerical spectrum with the dispersion relations Eqs.\ref{62d} and \ref{68}, we extract from them the expressions relating $\Re(\gamma)$ to $\Im(\gamma)$: 
\begin{eqnarray}
\label{69} \frac{\widehat{\gamma}_r}{\Theta} &\approx &\sqrt{2}/2(\widehat{\gamma}_i+1)(-\widehat{\gamma}_i)^{3/2}, \\
\label{70} \frac{\widehat{\gamma}_r}{\Theta} &\approx & (\sqrt{\eta\omega_*}/\rho^2) (\widehat{\gamma}_i+1)\widehat{\gamma}_i^2\left(-1-\rho^2k_y^2-\widehat{\gamma}^{-1}\right)^{1/2},
\end{eqnarray} 
where Eq.\ref{69} describes the even modes, Eq.\ref{70} the odd modes and $\Theta \equiv \pi (\sqrt{\eta\omega_*}/\rho^2)(\rho/L_x)$, $\widehat{\gamma}_{r}\equiv \Re(\gamma)/\omega_*$, $\widehat{\gamma}_{i}\equiv \Im(\gamma)/\omega_*$. In Fig.\ref{fig3} we compare the previous expressions with the numerical data for different values of $\sqrt{\eta\omega_*}/\rho^2$, finding excellent agreement. The small discrepancy around the maximum growth rate is due to the fact that the regimes investigated are not sufficiently asymptotic. 
We also tested the predictions of the dispersion relations Eqs.\ref{62d} and \ref{68} for the complex frequency of the modes (not just the shape of the spectrum), obtaining again a good match with the numerical results.  

\subsubsection{Finite $L_n/L_s$ corrections}\label{FbN}

In this Subsection, we discuss the effect of a finite $L_n/L_s$ on the stability of the system and we therefore complete the characterisation of the modes that was interrupted in Subsection \ref{FbT}. In the limit of small $\sqrt{\eta\omega_*}/\rho^2$, we identified numerically that the growth rate can be written as:
\begin{equation}
\label{71} \Re(\gamma)=\Re(\gamma_0)F\left(\frac{L_n}{L_s},k_y\rho,\frac{L_x}{\rho}\right),
\end{equation}
where $\gamma_0$ is the $L_n/L_s=0$ growth rate given by Eqs.\ref{62d} and \ref{68}. In other words, the effect of the resistivity and of $L_n/L_s$ are independent from each other  and $F$ is the correction function to be numerically characterised. In addition, it is worth noticing that, in the fluid limit, $k_y\rho$ has only a weak effect on the correction function $F$. Indeed, for our equations to be valid, $k_y\rho$ must be much smaller than unity and this parameter appears in Eqs.\ref{63a} and \ref{62a} only in $\kappa$ as $(1+k_y^2\rho^2)$ (see Eq.\ref{61b}). For modes with $k_y\rho \sim 1$ a kinetic treatment would be needed, but this is outside the scope of the present work. Note also that $F$ has an extra hidden parameter, the mode number $n$ and it is different for even and odd eigenmodes. In the numerical investigation we studied cases with $L_x/\rho=[5;10;15;20;25;30;35]$ and $\rho k_y=0.1$ (simulations performed with $\rho=0.5$ produced results that were very similar to those presented). We fixed $\sqrt{\eta\omega_*}/\rho^2=0.12$, but we checked that our results were not changing for smaller values of this parameter (we calculated $F$ for $\sqrt{\eta\omega_*}/\rho^2$ as small as $0.04$ without finding differences). 

\begin{figure}
\includegraphics[height=8cm]{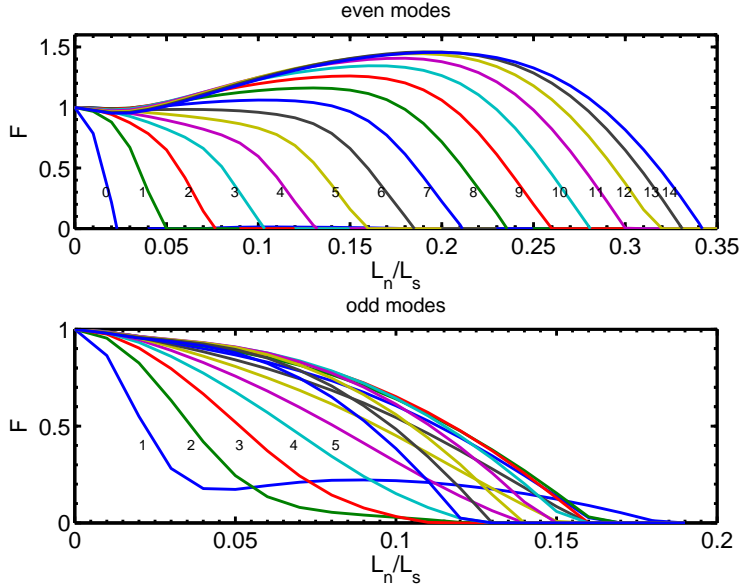} \caption{Correction function, $F$, as a function of $L_s/L_n$. Different mode numbers, $n$ are represented by different curves and labelled on the figure whenever possible. For this case, $L_x/\rho=15$, $k\rho=0.1$.}
\label{fig4}
\end{figure}  
Figure \ref{fig4} shows how $F$ varies for the first 15 even and odd modes as $L_n/L_s$ is increased and $L_x/\rho=15$. For both parities, the magnetic shear eventually stabilizes the modes. This is similar to what happens in the infinite systems (i.e. $L_x/\rho\rightarrow \infty$) studied in previous works, although in our case we observe complete suppression of the mode only when $L_n/L_s$ crosses a finite critical value. In addition, surprisingly, we find windows of $L_n/L_s$ in which some of the even modes can become more unstable (i.e. $F>1$). The amount of this shear induced destabilization can be significant and depends on the mode number, on the size of the system (i.e. $L_x/\rho$) and weakly on $\rho k_y$. It is also interesting to note that small $n$ even and odd modes can invert their stabilization trend and form a second unstable branch with relatively small growth rate (especially for the even modes) but less effected by the magnetic shear (see the odd $n=1$ mode in Fig.\ref{fig4}). 
\begin{figure}
\includegraphics[height=8cm]{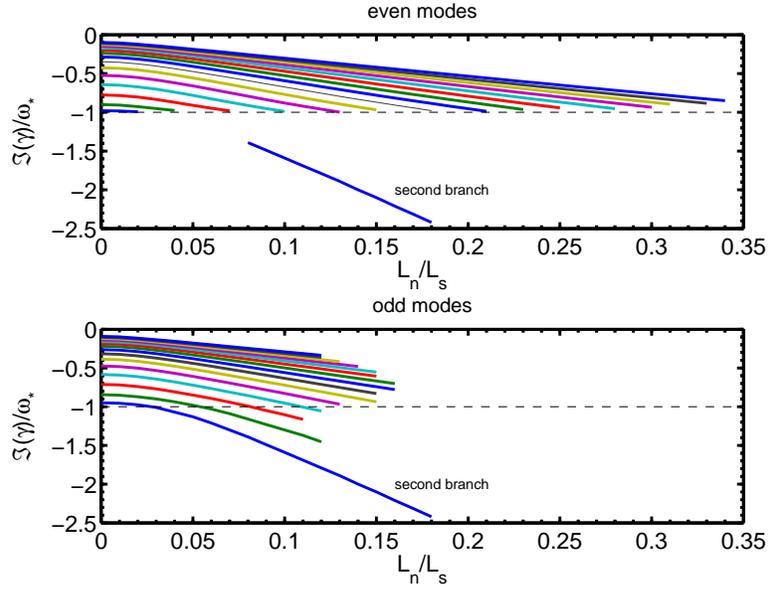} \caption{Rotation frequency as a function of $L_n/L_s$ for the the same parameters as Fig.\ref{fig4}. The mode numbers go from 0 to 14 for the even modes and from 1 to 15 for the odd and increase from the bottom curve to the upper.}
\label{fig4b}
\end{figure}  
In general, as $L_n/L_s$ is increased and the even modes approach their marginally stable state, their rotation frequency matches the diamagnetic frequency, as can be seen in Fig.\ref{fig4b}. On the other hand, the second unstable branch described above is associated to quickly rotating modes with $\Im(\gamma)>\omega_*$, and has therefore a different character. The situation is different for the odd modes, which do not stabilize at a specific rotation frequency. On the other hand, also the odd modes enter the second unstable branch when their frequency is higher than $|\omega_*|$.  

\begin{figure}
\includegraphics[height=8cm]{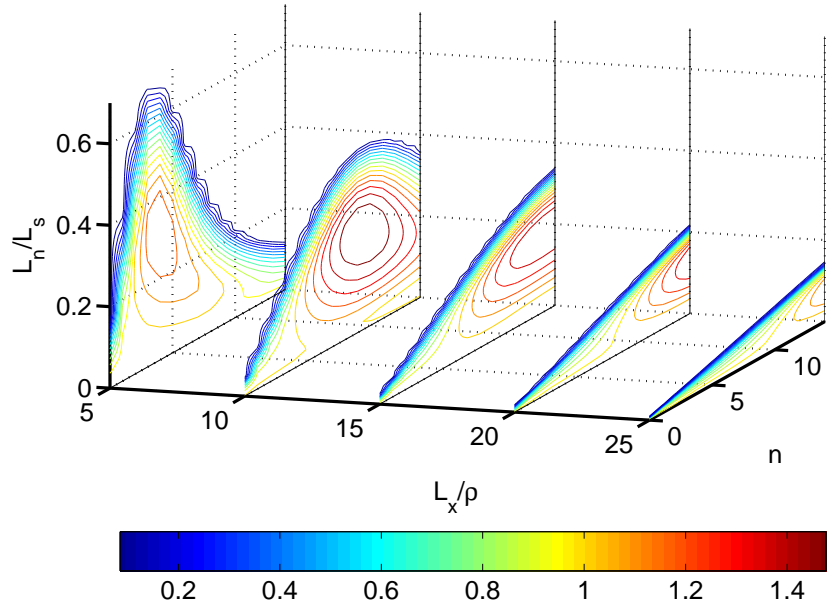} \caption{Contour plots of $F$ as a function of $L_s/L_n$ and $n$. The five slices correspond to different $L_x/\rho$.}
\label{fig6}
\end{figure}  
\begin{figure}
\includegraphics[height=8cm]{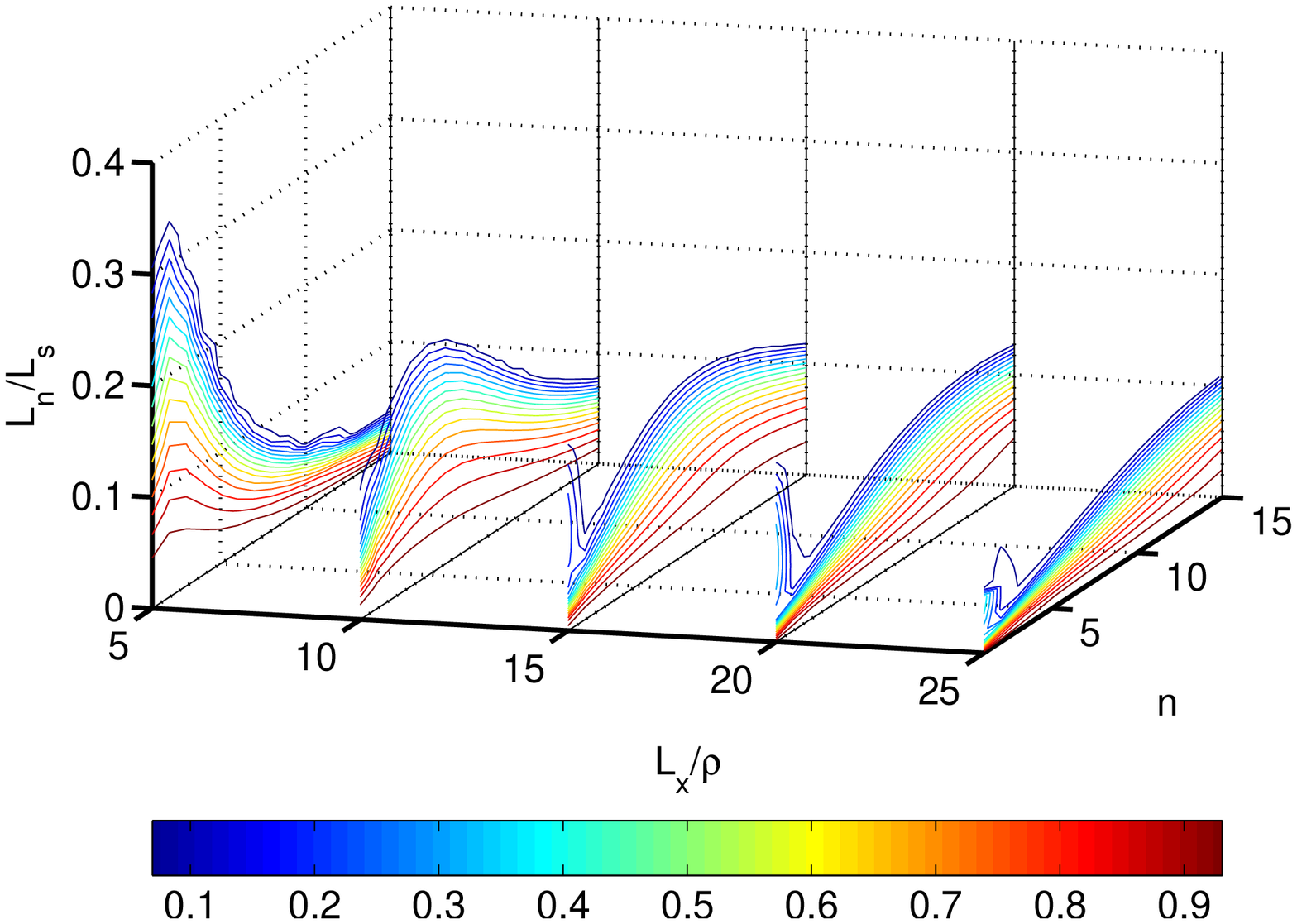} \caption{Same as Fig.\ref{fig6} for odd modes.}
\label{fig6b}
\end{figure}  
Figure \ref{fig6} and \ref{fig6b} show contour plots of $F$ for the even and odd modes as a function of $n\leq 15$ and $L_n/L_s$ for $k_y\rho=0.1$ and five values of $L_x/\rho$. Only contour levels for $F \geq 0$ are plotted, so that the upmost line in each slice marks the marginally stable conditions. These figures reveal that the shear stabilization is more efficient at larger $L_x/\rho$. Both small and large $n$ perturbations (i.e. small and large $k_x$) are more effectively damped than intermediate mode numbers. In addition, the mode numbers of the most resilient modes shift to larger values as $L_x/\rho$ increases. Note also that the windows of shear destabilization discussed above occur only for the even modes and that are more effective for $L_x/\rho\approx 10$ and less significant for smaller and larger values of $L_x/\rho$. The new branch of the odd modes is clearly displayed in the low $n$ region of Fig.\ref{fig6b}, where it becomes more prominent as $L_x/\rho$ is increased. 

\begin{figure}
\includegraphics[height=8cm]{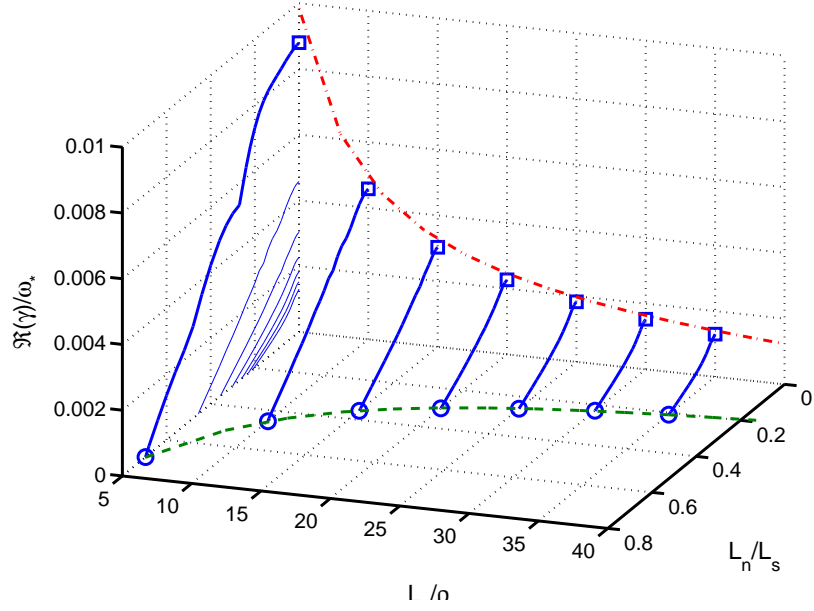} \caption{Maximum growth rate of the even modes normalized to $\omega_*$ as a function of $L_n/L_s$ and $L_x/\rho$. The squares represent the vales for $L_n/L_s=0$ and are compared with the theoretical curve shown as a dash-dotted line (see text). The circles mark the stabilization threshold and are interpolated by Eq.\ref{72}. All the curves at different $L_x/\rho$ are projected on the $L_x/\rho=5$ plane for easier comparison.}
\label{fig7}
\end{figure}  
\begin{figure}
\includegraphics[height=8cm]{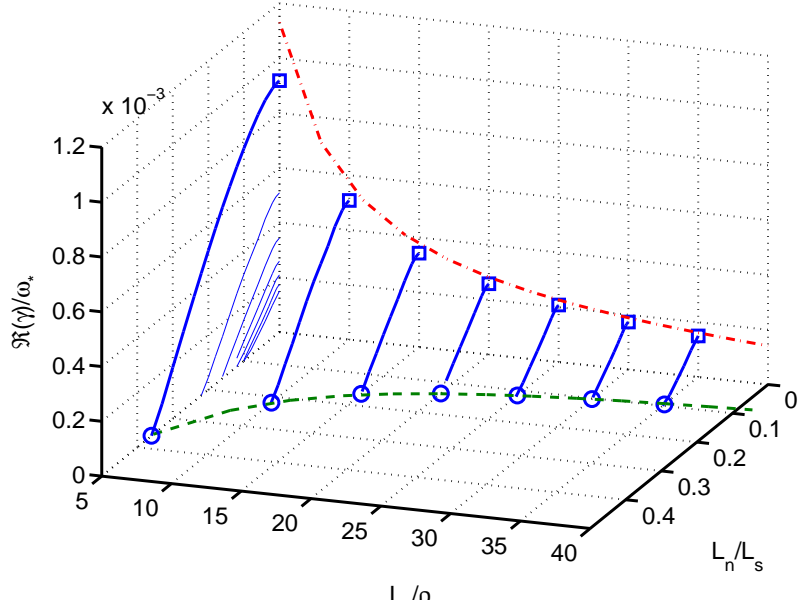} \caption{Same as Fig.\ref{fig7} for the odd modes. The circles are interpolated by Eq.\ref{73}}
\label{fig8}
\end{figure} 
It is interesting to determine the behaviour of the most unstable mode as a proxy of the overall stability of the equilibrium. In Figs.\ref{fig7} and \ref{fig8} we plot $\Re(\gamma)_{max}/\omega_*$ as a function of $L_n/L_s$ and $L_x/\rho$ for the even and odd modes. In the plane $L_n/L_s=0$ we have added a dashed line representing the theoretical predictions given at the end of Secs.\ref{Small_res} and \ref{Small_res_even}, finding in both cases a good agreement. For the even modes, the maximum growth rate shows an approximately linear decay as a function of $L_n/L_s$, while for the odd modes the damping is well represented by a cubic function. To facilitate the comparison between these curves, we duplicated them on the $L_x/\rho=5$ plane. 

The critical $L_n/L_s$ for the complete stabilization of the modes as a function of the system size is well represented with a power law decay. In particular, we find that our numerical results are well matched by:
\begin{equation}
\label{72} \left(\frac{L_n}{L_s}\right)_{cr} \approx 1.87 \left(\frac{L_x}{\rho}\right)^{-0.61},
\end{equation}
for the even modes and:
\begin{equation}
\label{73} \left(\frac{L_n}{L_s}\right)_{cr} \approx 1.08 \left(\frac{L_x}{\rho}\right)^{-0.69},
\end{equation}
for the odd modes. Note the relatively weak exponent in both expressions, suggesting that the critical value remains significant even in large systems. These two expressions are probably the most important result of the present work as they suggest that the unconditional drift wave stability obtained in \cite{Guzdar1978,Chen1980} is exclusive to infinite systems in which all the energy is dispersed through the boundary conditions. Equations \ref{72} and \ref{73} and the numerical data they represent are plotted in the $\Re(\gamma)_{max}=0$ plane of Figs.\ref{fig7} and \ref{fig8}. 

\section{Discussion}

In all the limits discussed, a combination of density gradients, finite resistivity and finite system size can destabilize a spectrum of unstable modes. The effect of the magnetic shear is generally stabilizing, although relatively small values of it can make some of the modes more unstable (this effect never occurs for the fastest growing mode). 

The destabilization is due to the fact that the reflecting boundary conditions that we are imposing trap the energy of the waves. The same effect would therefore occur in periodic configurations, which are often used in numerical simulations (see the Appendix). It is not uncommon to find in literature works that employ numerical domains of a few tens of $\rho$ and relatively small $L_n/L_s$, compatible with a significant spectrum of unstable modes. One example is \cite{Militello2008b} in which turbulence is driven by the nonlinear version of the modes presents here. Our work suggests that reflecting or periodic boundary conditions can destabilize or enhance the growth rate of the drift waves and therefore affect the anomalous transport estimated in numerical simulations. 

A real plasma, however, is an open system in which the energy is, in principle, not bounded to a specific region. Nevertheless, the destabilizing mechanism that we described could play a role in a number of realistic situations. For example, plasma inhomogeneities can reflect part of the energy back to its source (i.e. the resonant surface), similarly to what would happen to a pulse travelling in a rope consisting of two sections of different thickness. As the reflection would not be perfect, we expect that the calculations we presented would describe the worst case scenario (i.e. the highest limit for the growth rate). In addition, in toroidal systems, different poloidal modes are connected with each other through curvature coupling. This would allow modes that resonate at different positions to exchange energy so that an outgoing wave emitted on one resonant surface might look like an ingoing wave for a neighbouring surface, as originally proposed by Taylor \cite{Taylor1976}. Using this picture, our calculations in the simpler and more intuitive slab geometry helps to clarify the mechanism which destabilizes drift waves in toroidal configurations \cite{Chen1980}. In this case, $L_x$ would represent the distance between two resonant surfaces. 

It is important to notice that, for the instability to occur, the energy reflection must take place before the wave reaches the region where the ion Landau damping is strong. This region is characterized by the requirement that $k_\parallel v_{th,i}/\omega \sim 1$, where $k_\parallel =k_yx/L_s$ and $v_{th,i}=(T_i/m_i)^{1/2}$ is the ion thermal velocity (see, e.g. \cite{Perlstein1969}). The nature of the problem is therefore determined by the length scale $L_{Ld} \equiv L_s \omega/(k_y v_{th,i})$ and the condition $L_x<L_{Ld}$ is necessary for instability. Note that assuming $\omega \sim \omega_*$, we have that $L_{Ld} = \rho_s (L_s/L_n)\sqrt{T_e/T_i}$. In the particular case of cold ions treated in this paper $L_{Ld} \rightarrow \infty$, thus assuring that our calculation is consistent.

In our model, the role played by the resistivity is non trivial. Indeed, it is essential in order to drive the unstable modes and appears in the equations in the form of a singular perturbation. In this respect, the unstable drift waves we discussed are similar to the resistive tearing modes \cite{Furth1963}. On the other hand, the resistivity can also provide a dissipation mechanism and, if too large, it can even damp the perturbations. As noticed at the end of Section \ref{SecIVA}, in the $\beta=\rho=0$ regime the growth rate vanishes for both $\eta=0$ and $\eta\rightarrow \infty$. The damping effect becomes also visible in the case of infinite systems, such as those examined in \cite{Guzdar1978,Chen1980}, where the resistivity is stabilizing the modes.      

Our work is limited by the fact that we used a simple fluid model. Finite Larmor radius effects are restricted by the conditions $k\rho\ll 1$ and for $\rho k_x\approx (\rho/L_x)\pi n \ll 1$ so that the modes analysed are correctly described if their mode number is relatively small. Electron inertia is neglected, together with electron wave particle interactions. This prevents us from properly treating collisionless regimes which, within our model are stable but might become unstable upon reintroduction of these effects. The calculation is electrostatic an approximation that is not justified in the edge region of the fusion devices \cite{Militello2011}. Interesting physics might occur when tearing modes are coupled to the modes described in this paper. In particular, we expect density driven (not current driven) magnetic islands, resulting from the electromagnetic version of the odd parity modes (which would have even electromagnetic flux at the resonant surface). This $\Delta'$ independent modes might be related to the microtearing modes. This problem will be addressed and discussed in a subsequent article.   

\section{Summary and Conclusions}

We investigated the linear stability of finite size resistive inhomogeneous plasmas in a fluid approximation. The eigenmodes of the unstable drift waves, or universal instabilities, that we have analysed are standing waves generated by the reflective boundary conditions that are applied at a finite distance from the resonant surface. The study of the effect of these boundary conditions on the stability of the equilibrium was the main scope of the work presented. We found that the wave reflection provides a robust destabilization mechanism, which can persist also in relatively large systems. 

The general dispersion relation characterising the modes was given in Eq.\ref{61}. This expression, however, is difficult to interpret and requires simplifications in order to make its physics more transparent. Four dimensionless parameter, $\sqrt{\eta \omega_*}/\rho^2$, $L_n/L_s$, $k\rho$ and $L_x/\rho$, govern the problem and determine the complex frequency of the modes. 

Several limits of Eq.\ref{61} were investigated by exploiting the smallness of some of the above mentioned parameters. In particular, we were able to find exact analytic solutions in the regime $L_x/\rho\ll min(L_s/L_n,\overline{\gamma}^{1/2}\sqrt{\eta\omega_*}/\rho^2)$, which were reported in Eqs.\ref{31} and \ref{32} and in the regime $L_x/\rho\gg \overline{\gamma}^{1/2}\sqrt{\eta\omega_*}/\rho^2$, described in Eqs.\ref{35}. These simple studies allowed us to determine that the system can indeed be unstable when is bounded at a finite distance $L_x$ and that its drive mechanism is the density gradient, combined by a finite resistivity. It is important to notice that, the in the limit $L_x\rightarrow \infty$, Eq.\ref{61} correctly becomes the dispersion relation derived in \cite{Guzdar1978,Chen1980} for infinite systems with outgoing wave boundary conditions, which predicts stability. 

For small but finite values of $\sqrt{\eta\omega_*}/\rho^2$, Eq.\ref{61} becomes Eq.\ref{63a} and Eq.\ref{62a} for even and odd instabilities respectively. These expressions are still implicit in the complex frequency and the stabilization and destabilization mechanisms are difficult to identify. Finally, by taking their limit for small $L_n/L_s$ we arrived to Eq.\ref{62d} and Eq.\ref{68} which describe a spectrum of unstable odd and even modes. The important effect of the magnetic shear could not be cast in a simple analytic form (but it is rigorously contained in Eqs.\ref{63a} and \ref{62}) and we therefore resorted to a numerical characterisation (Sec.\ref{FbN}). In particular, we found that the growth rate of the most unstable mode is reduced by a finite magnetic shear, the critical value of which is finite and depends on the position of the wave reflection (see Eqs.\ref{72} and \ref{73}). This is probably our most important result as it shows that unconditional stability is a peculiar feature of the infinite systems and is not reproduced in more general configurations.

Under the condition that the at least partial wave reflection occurs before the mode can be damped by wave-particle interactions, we therefore expect to find unstable modes driven by density gradients in the presence of finite (albeit relatively small) magnetic shear. In other words, our conclusion is that the universal instability can exist in realistic plasmas and definitely in numerical simulations.  

\section{Acknowledgements}
F.M. acknowledges enlightening discussions with Dr. B. Taylor, J. Hastie and  Dr. J. Connor on the connection between the work presented and the full toroidal problem. This work was part-funded by the RCUK Energy Programme under grant EP/I501045 and the European Communities under the contract of Association between EURATOM and CCFE and between EURATOM and CEA. To obtain further information on the data and models underlying this paper please contact PublicationsManager@ccfe.ac.uk. The views and opinions expressed herein do not necessarily reflect those of the European Commission.

\appendix

\section{Periodic boundary conditions}

\begin{figure}
\includegraphics[height=8cm]{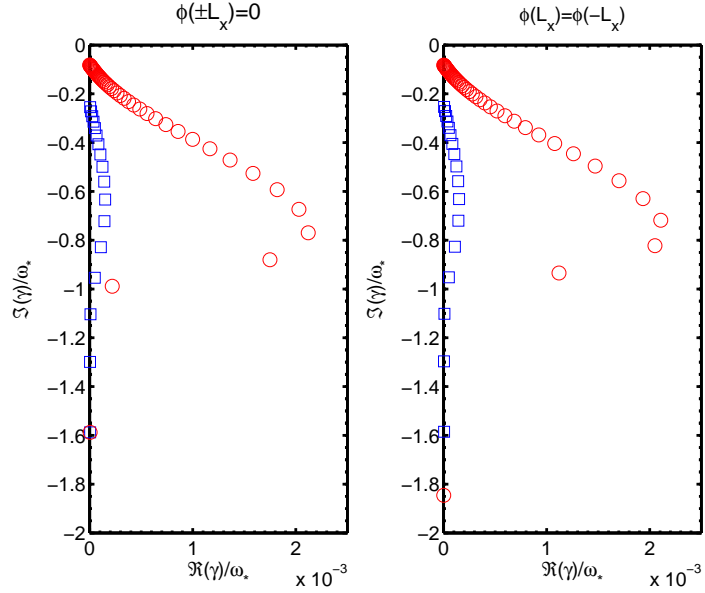} \caption{Spectrum of the modes for identical simulation parameters with Dirichlet (left panel) and periodic (right panel) boundary conditions. For both simulations  $\sqrt{\eta \omega_*}/\rho^2=0.12$, $L_n/L_s=0.1$, $k\rho=0.1$ and $L_x/\rho=15$. Circles represent even modes, while squares odd modes.}
\label{fig9}
\end{figure}

The extension of the calculation presented in the text to a case with periodic boundary conditions is relatively straightforward. In particular, the Dirichlet boundary conditions $\phi(\pm L_x)=0$ applied to Eq.\ref{6} should be replaced with $\phi(L_x)=\phi(-L_x)$. The nature of the problem would not change significantly as the energy trapping, responsible for the unstable modes, would persist with the new boundary conditions. Although we did not solve the analytical problem and obtained a dispersion relation for the periodic case, we performed several numerical simulations confirming that the unstable mechanism is present. As an example, we show in Fig.\ref{fig9} the unstable part of the spectrum for the same problem with Dirichelt and periodic boundary conditions (with $\sqrt{\eta \omega_*}/\rho^2=0.12$, $L_n/L_s=0.1$, $k\rho=0.1$ and $L_x/\rho=15$). 

\end{document}